\documentclass[prd,superscriptaddress,amsfonts,amssymb,amsmath,showpacs,twocolumn]{revtex4-2}
\usepackage{bm}
\usepackage{amsfonts}
\usepackage{latexsym}
\usepackage[latin1]{inputenc}
\usepackage{graphicx}
\usepackage{amsmath}
\usepackage{palatino}
\usepackage{mathpazo}
\usepackage{textcomp}
\usepackage{float}
\usepackage{booktabs}
\usepackage{dcolumn}
\usepackage{booktabs}
\usepackage{multirow}
\usepackage{hyperref}
\hypersetup{colorlinks,citecolor=blue}
\usepackage{amsmath}
\usepackage{xcolor}
\usepackage{orcidlink}
\usepackage[caption=false]{subfig}
\usepackage{commath}
\def\jnl@style{\it}
\def\aaref@jnl#1{{\jnl@style#1}}

\def\aaref@jnl#1{{\jnl@style#1}}

\def\aj{\aaref@jnl{AJ}}                   
\def\apj{\aaref@jnl{ApJ}}                 
\def\apjl{\aaref@jnl{ApJ}}                
\def\apjs{\aaref@jnl{ApJS}}               
\def\apss{\aaref@jnl{Ap\&SS}}             
\def\aap{\aaref@jnl{A\&A}}                
\def\aapr{\aaref@jnl{A\&A~Rev.}}          
\def\aaps{\aaref@jnl{A\&AS}}              
\def\mnras{\aaref@jnl{Mon.~Not.~Roy.~Astron.~Soc.}}             
\def\prd{\aaref@jnl{Phys.~Rev.~D}}        
\def\prc{\aaref@jnl{Phys.~Rev.~C}}  
\def\prl{\aaref@jnl{Phys.~Rev.~Lett.}}    
\def\qjras{\aaref@jnl{QJRAS}}             
\def\skytel{\aaref@jnl{S\&T}}             
\def\ssr{\aaref@jnl{Space~Sci.~Rev.}}     
\def\zap{\aaref@jnl{ZAp}}                 
\def\nat{\aaref@jnl{Nature}}              
\def\aplett{\aaref@jnl{Astrophys.~Lett.}} 
\def\apspr{\aaref@jnl{Astrophys.~Space~Phys.~Res.}} 
\def\physrep{\aaref@jnl{Phys.~Rep.}}      
\def\physscr{\aaref@jnl{Phys.~Scr}}       
\def\commat{\aaref@jnl{Comm.~Math.~Phys.}}              
\def\science{\aaref@jnl{Science}}               
\def\cqg{\aaref@jnl{Classical Quant.~Grav.}}            
\def\jpcs{\aaref@jnl{JPCS}}                                     
\def\ijmpd{\aaref@jnl{Int.~J.~Mod.~Phys.~D}}                    
\def\grg{\aaref@jnl{Gen.~Relat.~Gravit.}}               
\def\rpp{\aaref@jnl{Rep.~Prog.~Phys.}}          
\def\npa{\aaref@jnl{Nucl.~Phys.~A}}        
\def\lrr{\aaref@jnl{Living Rev.~Rel.}}                   
\def\jcap{\aaref@jnl{J.~Cosmology Astropart.~Phys.}}    
\def\rmp{\aaref@jnl{Rev.~Mod.~Phys.}}   
\def\epjc{\aaref@jnl{Eur.~Phys.~J.~C}}

\allowdisplaybreaks[1]

\begin{document}

\color{black}       

\title{Bulk viscous fluid in symmetric teleparallel cosmology: theory versus experiment}

\author{Raja Solanki\orcidlink{0000-0001-8849-7688}}
\email{rajasolanki8268@gmail.com}
\affiliation{Department of Mathematics, Birla Institute of Technology and
Science-Pilani,\\ Hyderabad Campus, Hyderabad-500078, India.}
\author{Simran Arora\orcidlink{0000-0003-0326-8945}}
\email{dawrasimran27@gmail.com}
\affiliation{Department of Mathematics, Birla Institute of Technology and
Science-Pilani,\\ Hyderabad Campus, Hyderabad-500078, India.}
\author{P.K. Sahoo\orcidlink{0000-0003-2130-8832}}
\email{pksahoo@hyderabad.bits-pilani.ac.in}
\affiliation{Department of Mathematics, Birla Institute of Technology and
Science-Pilani,\\ Hyderabad Campus, Hyderabad-500078, India.}
\author{P.H.R.S. Moraes}
\email{moraes.phrs@gmail.com}
\affiliation{Universidade Federal do ABC (UFABC)-Centro de Ci\^encias Naturais e Humans (CCNH)-Avenida dos Estados 5001, 09210-580, Santo Andr\'e, SP, Brazil.}

\date{\today}

\begin{abstract}
The standard formulation of General Relativity Theory, in the absence of a cosmological constant, is unable to explain the responsible mechanism for the observed late-time cosmic acceleration. On the other hand, by inserting the cosmological constant in Einstein's field equations it is possible to describe the cosmic acceleration, but the cosmological constant suffers from an unprecedented fine-tunning problem. This motivates one to modify Einstein's space-time geometry of General Relativity. The $f(Q)$ modified theory of gravity is an alternative theory to General Relativity, where the non-metricity scalar $Q$ is the responsible candidate for gravitational interactions. In the present work we consider a Friedmann-Lem\^aitre-Robertson-Walker cosmological model dominated by bulk viscous cosmic fluid in $f(Q)$ gravity with the functional form $f(Q)=\alpha Q^n$, where $\alpha$ and $n$ are free  parameters of the model. We constrain our model with the recent Pantheon supernovae data set of 1048 data points, Hubble data set of 31 data points and baryon acoustic oscillations data set consisting of six points. For higher values of redshift, it is clear that the $f(Q)$ cosmology better fits data than standard cosmology. We present the evolution of our deceleration parameter with redshift and it properly predicts a transition from decelerated to accelerated phases of the universe expansion. Also, we present the evolution of density, bulk viscous pressure and the effective equation of state parameter with redshift. Those show that bulk viscosity in a cosmic fluid is a valid candidate to acquire the negative pressure to drive the cosmic expansion efficiently.We also examine the behavior of different energy conditions to test the viability of our cosmological $f(Q)$ model. Furthermore, the statefinder diagnostics are also investigated in order to distinguish among different dark energy models.  
\end{abstract}

\maketitle

\section{Introduction}\label{sec1}

The acceleration of the universe expansion is one of the most active discoveries of modern cosmology. Observational studies include Type Ia supernovae \cite{Riess,Perlmutter}, large scale structure \cite{T.Koivisto,S.F.}, baryon acoustic oscillations \cite{D.J.,W.J.} and cosmic microwave background radiation \cite{R.R.,Z.Y.}. The reason behind the late-time acceleration is a mystery. Several models hypothesize the existence of a component called {\it dark energy}, which makes up around 70\% of the entire universe and could possess the feature of speeding up the universe expansion. 

The cosmological constant $\Lambda$ in General Relativity (GR) field equations plays the role of dark energy, i.e., a fluid with constant energy density and high negative pressure. There are some issues with the cosmological constant model, like the cosmic coincidence problem \cite{dalal/2001}, which is the fact that the density of non-relativistic matter and dark energy are the same order today. A more delicate issue surrounding the cosmological constant is the so-called {\it cosmological constant problem}, which is the high discrepancy between the astronomically observed value of $\Lambda$ \cite{Riess,Perlmutter} and the Particle Physics theoretically predicted value of the quantum vacuum energy \cite{weinberg/1989}.

With the main purpose of solving the above cosmological issues, dynamical (time-varying) dark energy models such as Chaplygin gas model \cite{M.C.,A.Y.}, k-essence \cite{T.Chiba,C.Arm.}, quintessence \cite{Carroll,Y.Fujii} and decaying vacuum models \cite{xu/2011,tong/2011,freese/1987,abdel-rahman/1992} have been proposed for some time in the literature. 

Modified theories of gravity have also been intensively investigated to understand the origin of the cosmic acceleration as well as to address the cosmological constant model problems. It is possible to predict late-time cosmic acceleration by modifying GR action. Some possibilities can be seen within $f(R)$ \cite{appleby/2007,amendola/2007,saffari/2008}, $f(G)$ \cite{cognola/2006,li/2007}, $f(R,\mathcal{T})$ \cite{moraes/2016} and $f(T)$ \cite{ren/2021,nashed/2015,setare/2013} theories of gravitation, with $R$, $G$, $\mathcal{T}$ and $T$ being respectively the Ricci, Gauss-Bonnet, energy-momentum and torsion scalars. 

In the present article we will work with the recently introduced $f(Q)$ theory of gravity \cite{J.B.}, for which $Q$ is the non-metricity scalar, to be presented below. The non-metricity formulation has been discussed earlier by Hehl and Ne'eman (see References \cite{H1,H2,H3,H4}). The symmetric teleparallel gravity is proven in the so-called coincidence gauge by imposing that the connection is symmetric \cite{nester/1999}. Weyl geometry is also observed to be a particular example of Weyl Cartan geometry in which torsion disappears. The non-metricity is interpreted as a massless spin 3-field in the case of symmetric connections \cite{N1,N2}. Also, it is noted in the literature that due to the appearance of non-metricity, the light cone structure is not preserved during parallel transport \cite{H5}. Further, fermions are an issue in TEGR because they couple to the axial contorsion of the Weitzenbock connection. This difficulty is eliminated in STEGR since Dirac fermions only couple to the completely antisymmetric component of the affine connection and is unaffected by any disformation piece.


Although recently proposed, the $f(Q)$ gravity theory already presents some interesting and valuable applications in the literature \cite{Noemi/2021,Barros/2020,Ferreira/2022}. The first cosmological solutions in $f(Q)$ gravity appear in References \cite{jimenez/2020,khyllep/2021}, while $f(Q)$ cosmography and energy conditions can respectively be seen in \cite{mandal/2020,mandal/2020b}.

Here we are going to consider $f(Q)$ cosmology in the presence of a viscous fluid. When a cosmic fluid expands too fast, the recovering of thermodynamic equilibrium generates an effective pressure. The high viscosity in a cosmic fluid is the manifestation of such an effective pressure \cite{J.R.,H.O.}. 

Basically, there are two viscosity coefficients, namely shear viscosity and bulk viscosity. Shear viscosity is related to velocity gradients in the fluid, and by considering the universe as described by homogeneous and isotropic Friedmann-Lem\^aitre-Robertson-Walker (FLRW) background, it can be omitted. Anyhow, by dropping the FLRW background assumption, several cosmological models with shear viscosity fluid have been constructed, as one can check, for instance, \cite{bali/1988,bali/1987,deng/1991,huang/1990}. On the other hand, bulk viscosity, which we are going to consider here, introduces damping associated with volumetric straining. To get in touch with bulk viscous fluid cosmological models, one can check References \cite{samanta/2017,satish/2016,beesham/1993,colistete/2007,Davood/2019}. Moreover, some interesting applications of bulk viscous cosmology in black holes presented in \cite{bb1,bb2}.

Researchers examine dark energy reconstruction with numerous observations as data increases. The majority of studies has been concentrated on observable evidences from Type Ia supernovae, cosmic microwave background and baryon acoustic oscillations (BAO), which are known to be helpful in constraining cosmological models. The Hubble parameter dataset shows the intricate structure of the expansion of the universe. The ages of the most massive and slowly evolving galaxies offer direct measurements of the Hubble parameter $H(z)$ at various redshifts $z$, resulting in the development of a new form of standard cosmological probe \cite{Jimenez/2002}. 

In our present work we include 31 measurements of Hubble expansion spanned using differential age method \cite{Sharov/2017} and BAO data consisting of six points \cite{Blake/2011}. Scolnic et al. recently published a large Type Ia supernovae sample named Pantheon, with 1048 points and covering the redshift range $0.01< z< 2.3$ \cite{Scolnic/2018}. Our analysis uses the $H(z)$, BAO and Pantheon samples to constrain the cosmological model.

This work aims to describe the recently observed late-time acceleration with the help of bulk viscosity of cosmic fluid (without including any dark energy component) in the framework of  $f(Q)$ theory of gravity. The manuscript is organized as follows: in Sec. \ref{sec2} we discuss the $f(Q)$ gravity formalism. In Sec. \ref{sec3} we describe the FLRW universe dominated by bulk viscous non-relativistic matter and derive the expression for the Hubble parameter and the deceleration parameter. Further, in Sec. \ref{sec4}, we analyze the observational data to find the best-fit ranges for the parameters using the Hubble data set containing 31 points, BAO sample and the Pantheon data set of 1048 samples. Moreover, we analyze the behavior of different cosmological parameters such as Hubble, density, effective pressure, deceleration parameter and effective equation of state (EoS) parameter. In Sec. \ref{sec5}, we investigate the consistency of our bulk viscous fluid model by analyzing the different energy conditions. In Sec. \ref{sec6}, we analyze the behavior of statefinder parameters on the values constrained by the observational data to differentiate between dark energy models. Lastly, we discuss our results in Sec. \ref{sec7}.

\section{Fundamental formulations in $f(Q)$ gravity}\label{sec2}

In $f(Q)$ gravity theory, the spacetime is established with the help of non-metricity and symmetric teleparallelism condition, i.e. $\nabla_\alpha g_{\mu\nu} \neq 0$ and $R^\rho_{\sigma\mu\nu}=0$. The associated affine connection is given by


\begin{equation}\label{fq1}
	\Upsilon^\alpha_{\ \mu\nu}=\Gamma^\alpha_{\ \mu\nu}+L^\alpha_{\ \mu\nu},
\end{equation}
with 

\begin{eqnarray}
	\Gamma^\alpha_{\ \mu\nu}\equiv\frac{1}{2}g^{\alpha\lambda}(g_{\mu\lambda,\nu}+g_{\lambda\nu,\mu}-g_{\mu\nu,\lambda}),\label{fq2}\\
	L^\alpha_{\ \mu\nu}\equiv\frac{1}{2}(Q^{\alpha}_{\ \mu\nu}-Q_{\mu \ \nu}^{\ \alpha}-Q_{\nu \ \mu}^{\ \alpha}),\label{fq3}
\end{eqnarray}
being the Christoffel symbols and the distortion tensor, respectively, where

\begin{eqnarray}
	Q_{\alpha\mu\nu}\equiv\nabla_\alpha g_{\mu\nu},\label{fq4}
\end{eqnarray}
is the non-metricity tensors. 


The non-metricity tensor given by Eq.\eqref{fq4} has following two traces,

\begin{equation}\label{2a}
Q_\alpha = Q_\alpha\:^\mu\:_\mu, 
\end{equation}

\begin{equation}\label{2b}
	\tilde{Q}_\alpha = Q^\mu\:_{\alpha\mu}.
\end{equation}

In addition, the superpotential tensor or non-metricity conjugate is given by

\begin{eqnarray}\label{2c}
4P^\lambda\:_{\mu\nu} &=& -Q^\lambda_{\ \mu\nu}+Q_{\mu\ \nu}^{\ \lambda}+Q_{\nu\ \mu}^{\ \lambda}+(Q^\lambda-\tilde{Q}^\lambda)g_{\mu\nu}\nonumber\\
&-&\frac{1}{2}(\delta^\lambda_{\ \mu}Q_\nu+\delta^\lambda_{\ \nu}Q_\mu).
\end{eqnarray}

Then the trace of the non-metricity tensor can be acquired as \cite{Ruth}

\begin{equation}\label{2d}
Q = -Q_{\lambda\mu\nu}P^{\lambda\mu\nu}. 
\end{equation}

The definition of the energy-momentum tensor for matter is 
\begin{equation}\label{2e}
\mathcal{T}_{\mu\nu} = \frac{-2}{\sqrt{-g}} \frac{\delta(\sqrt{-g}L_m)}{\delta g^{\mu\nu}}.
\end{equation}

Furthermore, one can obtain the following relation between the curvature tensor $R^\rho_{\sigma\mu\nu}$ and $\mathring{R}^\rho_{\sigma\mu\nu}$ corresponding to the connection $\Upsilon$  and $\Gamma$ as

\begin{equation}\label{2f}
R^\rho_{\sigma\mu\nu} = \mathring{R}^\rho_{\sigma\mu\nu} + \mathring{\nabla}_\mu L^\rho_{\nu\sigma} - \mathring{\nabla}_\nu L^\rho_{\mu\sigma} + L^\rho_{\mu\lambda} L^\lambda_{\nu\sigma} - L^\rho_{\nu\lambda} L^\lambda_{\mu\sigma} 
\end{equation}

and so

\begin{equation}\label{2g}
R_{\sigma\nu} = \mathring{R}_{\sigma\nu} + \frac{1}{2} \mathring{\nabla}_\nu Q_\sigma + \mathring{\nabla}_\rho L^\rho_{\nu\sigma} - \frac{1}{2} Q_\lambda L^\lambda_{\nu\sigma} - L^\rho_{\sigma\lambda} L^\lambda_{\rho\sigma} 
\end{equation}

\begin{equation}\label{2h}
R = \mathring{R} + \mathring{\nabla}_\lambda Q^\lambda - \mathring{\nabla}_\lambda \tilde{Q}^\lambda - \frac{1}{4} Q_\lambda Q^\lambda + \frac{1}{2} Q_\lambda \tilde{Q}^\lambda - L_{\rho\nu\lambda} L^{\lambda\rho\nu}
\end{equation}

Now, the connection \eqref{fq1} can be parameterized as \cite{J.B.}
 
\begin{equation}
\Upsilon^\alpha \,_{\mu \beta} = \frac{\partial x^\alpha}{\partial \xi^\rho} \partial_ \mu \partial_\beta \xi^\rho.
\end{equation}
Here, $\xi^\alpha = \xi^\alpha (x^\mu)$ is an invertible relation. Hence, it is always possible to find a coordinate system so that the connection $ \Upsilon^\alpha_{\ \mu\nu} $ vanishes. This situation is called coincident gauge and the covariant derivative $\nabla_{\alpha}$ reduces to the partial one $\partial_{\alpha} $. But in any other coordinate system in which this affine connection does not vanish, the metric evolution will be affected and result in a completely different theory \cite{AP,Avv}. Thus in the coincident gauge coordinate , we have
\begin{equation}\label{2i}
Q_{\alpha\mu\nu} = \partial_\alpha g_{\mu\nu}
\end{equation}
while in an arbitrary coordinate system,
\begin{equation}\label{2j}
Q_{\alpha\mu\nu}= \partial_\alpha g_{\mu\nu} - 2 \Upsilon^\lambda_{\alpha (\mu} g_{\nu)\lambda}. 
\end{equation}

Locally, GR does not distinguish between gravitational and inertial effects, however, by invoking frame fields, it is possible to covariantly define gravitational energy in the teleparallel approach \cite{maluf/2013}. The canonical frame is identified in the absence of both curvature and torsion and the canonical coordinates in the absence of inertial effects. New physics can emerge from such a formalism. Symmetric teleparallel gravity is broadly derived in three gravity theories based on the coordinate transformations \cite{Hohmann}. In this case, we use the spatially flat case, i.e. $f(Q)$ gravity, whose field equations are much easier to understand.

The $f(Q)$ gravity is described by the action \cite{J.B.}:
\begin{equation}\label{2k}
S= \int{\frac{1}{2}f(Q)\sqrt{-g}d^4x} + \int{L_m\sqrt{-g}d^4x},
\end{equation}
in which $f(Q)$ is an arbitrary function of the non-metricity, $g=\text{det}\  g_{\mu\nu}$ and $L_m$ is the lagrangian density of matter. One can check in Reference \cite{J.B.} that the  functional form $f(Q)=-Q $ corresponds to the STEGR (symmetric teleparallel equivalent to General Relativity) limit.

The gravitational field equations obtained by varying action \eqref{2k} with respect to the metric are
\begin{widetext}
\begin{equation}\label{2l}
\frac{2}{\sqrt{-g}}\nabla_\lambda (\sqrt{-g}f_Q P^\lambda\:_{\mu\nu}) + \frac{1}{2}g_{\mu\nu}f+f_Q(P_{\mu\lambda\beta}Q_\nu\:^{\lambda\beta} - 2Q_{\lambda\beta\mu}P^{\lambda\beta}\:_\nu) = -\mathcal{T}_{\mu\nu},
\end{equation}
\end{widetext}
in which we defined $ f_Q =df/dQ $.

Moreover, by varying the action with respect to the connection, one obtains

\begin{equation}\label{2m}
\nabla_\mu \nabla_\nu (\sqrt{-g}f_Q P^{\mu\nu}\:_\lambda) =  0.
\end{equation}

\section{The cosmological model}\label{sec3}

We consider the flat FLRW metric for our analysis \cite{SVV}, such that:
\begin{equation}\label{3a}
ds^2= -dt^2 + a^2(t)(dx^2+dy^2+dz^2).
\end{equation}
In Equation (\ref{3a}) above, $ a(t) $ is the cosmic scale factor, as usually. 

From now, we will fix the coincident gauge so that connection becomes trivial and the metric is only a fundamental variable. 

We are going to assume a bulk viscous fluid and below we present some considerations favoring such an assumption.

Firstly, to consider bulk viscosity in a fluid can be seen as an attempt to refine its  description, minimizing its ideal properties. This can be checked, for instance, in the stellar astrophysics realistic models in References  \cite{gusakov/2007,gusakov/2008,haensel/2002}.

Under conditions of spatial homogeneity and isotropy (which refers to the cosmological principle, as one can check, for instance, Reference \cite{ryden/2003}), the bulk viscous pressure is the unique admissible dissipative phenomenon. In a gas dynamical model, the existence of an effective bulk pressure can be traced back to a non-standard self-interacting force on the particles of the gas \cite{colistete/2007}. The bulk viscosity contributes negatively to the total pressure, as one can chegk, for instance References \cite{odintsov/2020,fabris/2006,meng/2009}.

Due to spatial isotropy, the bulk viscous pressure is the same in all spatial directions and hence proportional to the volume expansion $\theta=3H$, with $H=\dot{a}/a$ being the Hubble parameter and a dot represents time derivative.

The effective pressure of the cosmic fluid becomes \cite{brevik/2005,gron/1990,C.E./1940}
\begin{equation}
\bar{p}= p-\zeta\theta=p-3\zeta H,
\end{equation}
in which $p$ is the usual pressure and $ \zeta >0 $ is the bulk viscosity coefficient, which we will assume as a free parameter of the model.

The corresponding energy-momentum tensor is given by 
\begin{equation}\label{3c}
 \mathcal{T}_{\mu\nu} = (\rho+\bar{p})u_\mu u_\nu+ \bar{p}g_{\mu\nu}, 
\end{equation}
in which $\rho$ is the matter-energy density and the four-velocity $ u^\mu $ is such that its components are $ u^\mu = (1,0,0,0)$. 

The relation between normal pressure and matter-energy density follows \cite{J/2006} $p=(\gamma-1)\rho$, with $\gamma$ being a constant lying in the range $0\leq \gamma \leq 2$. Then the effective equation of state for the bulk viscous fluid is given by the following:
\begin{equation}\label{3i}
\bar{p}= (\gamma -1)\rho - 3\zeta H.
\end{equation}

The Friedmann-like equations for our $f(Q)$ gravitational model are obtained from the substitution of Equations (\ref{3a})-(\ref{3i}) into Equation (\ref{2g}) and read  as follows (check, for instance, References \cite{Ruth,T/2018})
\begin{equation}\label{3d}
3H^2= \frac{1}{2f_Q}\left(-\rho + \frac{f}{2}\right),
\end{equation}
\begin{equation}\label{3e}
\dot{H}+\left(3H+\frac{\dot{f_Q}}{f_Q}\right)H = \frac{1}{2f_Q} \left(\bar{p}+\frac{f}{2}\right).
\end{equation}  
In particular, for $f(Q)=-Q$ we retrieve the usual Friedmann equations \cite{T/2018}, as expected, since as we have mentioned above, this particular choice for the functional form of the function $f(Q)$ is the STEGR limit of the theory.

For our investigation of bulk viscosity fluid cosmological model, we consider the following $f(Q)$ functional form: 
\begin{equation}\label{3f}
f(Q)= \alpha Q^n,  
\end{equation} 
with $\alpha\neq0 $ and constant $n$. This particular functional form for $f(Q)$ was motivated by a polinomial form applied, for instance, in Reference \cite{mandal/2020b}.

For the above choice of the $f(Q)$ function (Equation (\ref{3f})), we rewrite Equations (\ref{3d})-(\ref{3e}) as follows

\begin{equation}\label{3g}
\rho = \alpha6^n \left(\frac{1}{2}-n \right) H^{2n},
\end{equation}
\begin{equation}\label{3h}
\dot{H}+\frac{3}{2n} H^2= \frac{6^{1-n}\bar{p}}{2\alpha n(2n-1)} H^{2(1-n)},
\end{equation}  
in which for the former we have isolated $\rho$.

From Equation \eqref{3h} and Equation \eqref{3i}, we have the following:
 
\begin{equation}\label{3j}
\dot{H}+ \frac{3\gamma}{2n} H^2 = - \frac{6^{2-n}\zeta}{4\alpha n(2n-1)} H^{3-2n}. 
\end{equation} 

Now, we replace the term $d/dt$ by $d/dln a$ via the expression $d/dt=Hd/dln a$, such that Equation \eqref{3j} becomes
\begin{equation}\label{3k}
\frac{dH}{dln a} + \frac{3\gamma}{2n}H = -\frac{6^{2-n}\zeta}{4\alpha n(2n-1) } H^{2(1-n)}.
\end{equation}

The integration of Equation \eqref{3k} yields the following solution

\begin{equation}\label{3l}
H(a) = \left\{( H_0 a^{-\frac{3\gamma}{2n}} )^{2n-1} + \frac{6^{1-n}\zeta}{\gamma \alpha (2n-1)^2}[ a^{-\frac{3\gamma(2n-1)}{2n}}-1 ] \right\}^{\frac{1}{2n-1}},
\end{equation}
 with $H_0$ being a constant of integration to be found below.

We obtain the Hubble parameter in terms of redshift by using relation \cite{ryden/2003} $ a(t)= 1/(1+z)$ in Equation \eqref{3m}.  By making $z=0$ in (\ref{3m}) we find that $H(0)=H_0$. The deceleration parameter is defined as $q= -\ddot{a}a/\dot{a}^2 = -\ddot{a}/(H^2a) $. Henceforth from Equation \eqref{3l} we have
\begin{widetext}
\begin{eqnarray}\label{3m}
H(z) &=& \left\{[ H_0 (1+z)^{\frac{3\gamma}{2n}} ]^{2n-1} + \frac{6^{1-n}\zeta}{\gamma \alpha (2n-1)^2} [ (1+z)^{\frac{3\gamma(2n-1)}{2n}}-1 ] \right\}^{\frac{1}{2n-1}},\\
\label{q}
q(z) &=&\frac{3}{2n}\left\{\frac{\zeta}{\alpha 6^{n-1}(2n-1) \left\{ [ H_0(1+z)^{\frac{3\gamma}{2n}} ]^{2n-1} + \frac{6^{1-n}\zeta}{\gamma\alpha (2n-1)^2} [ (1+z)^{\frac{3\gamma(2n-1)}{2n}} -1 ] \right\}} + \gamma\right\} -1. 
\end{eqnarray}
\end{widetext}

\section{Observational constraints}\label{sec4}

To examine the observational features of our cosmological model, we use the most recent cosmic Hubble and Supernovae observations. We use 31 points of the Hubble data sets, 6 points of the BAO data sets and 1048 points from the Pantheon supernovae samples. We apply the Bayesian analysis and likelihood function along with the Markov Chain Monte Carlo (MCMC) method in \texttt{emcee} python library \cite{Mackey/2013}. 

\subsection{Hubble datasets}
 
The Hubble parameter can be expressed as $H(z)=-dz/[dt(1+z)]$. As $dz$ is derived from a spectroscopic survey, the model-independent value of the Hubble parameter may be calculated by measuring the quantity $dt$. 

{We incorporate the set of 31 data points that are measured from the differential age approach \cite{Sharov/2018} to avoid extra correlation with BAO data. The mean values of the model parameters $\zeta$, $\alpha$, $\gamma$ and $n$ are calculated using the chi-square function  as follows:

\begin{equation}
\chi _{H}^{2}(\zeta,\alpha,\gamma,n)=\sum\limits_{k=1}^{31}
\frac{[H_{th}(\zeta,\alpha,\gamma,n, z_{k})-H_{obs}(z_{k})]^{2}}{
\sigma _{H(z_{k})}^{2}}.  \label{4a}
\end{equation}

Here, $H_{th}$ is the Hubble parameter value predicted by the model, $H_{obs}$ represents its observed value and the standard error in the observed value of $H$ is $\sigma _{H(z_{k})}$. 

From the Hubble dataset, we obtain the best fit values for $\zeta$, $\alpha$, $\gamma$, $n$ as the $1-\sigma$ and $2-\sigma$ contour plots in Fig.\ref{ContourHub}. 

\begin{widetext}

\begin{figure}[H]
\centering
\includegraphics[scale=0.9]{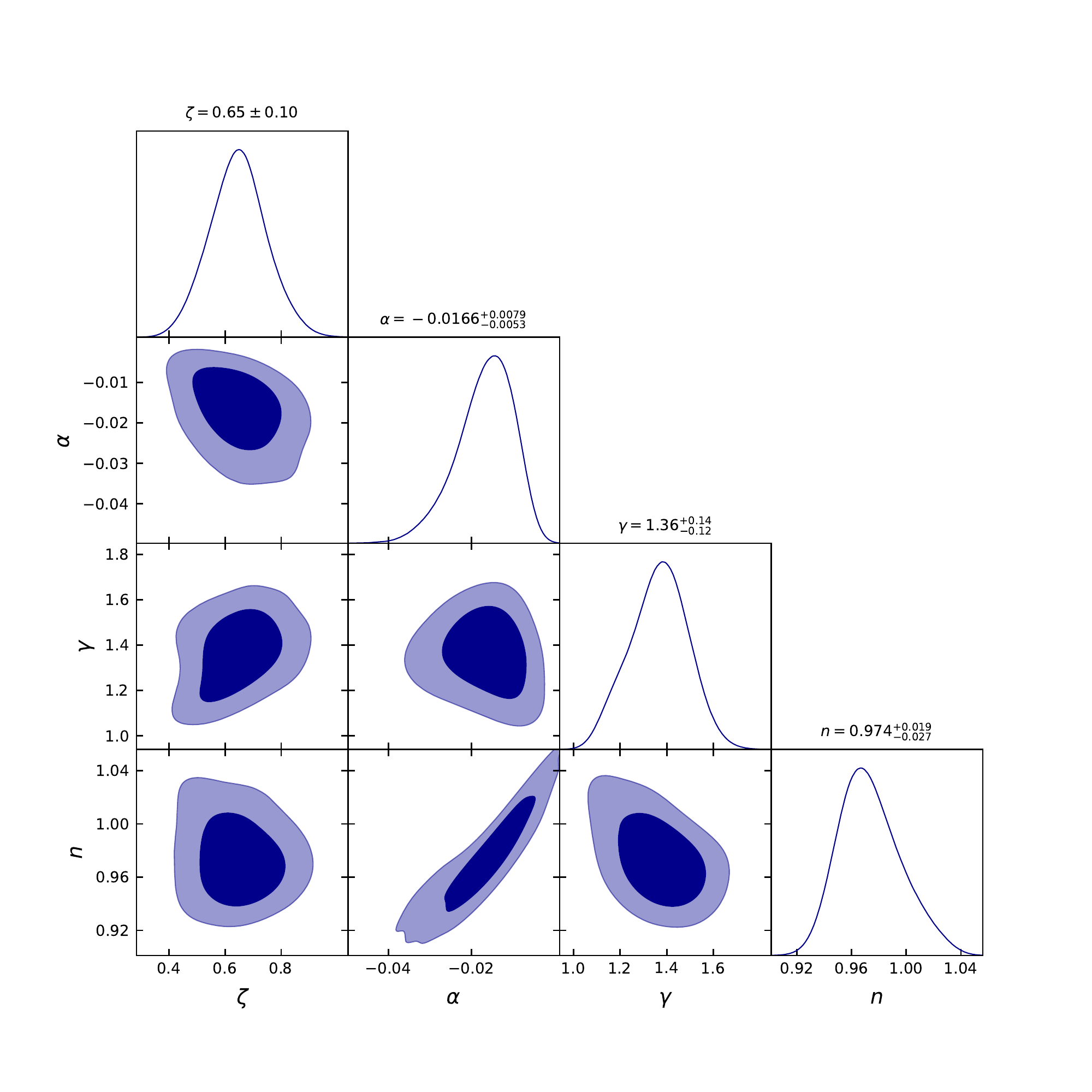}
\caption{The $1-\sigma$ and $2-\sigma$ likelihood contours for the model parameters using the Hubble datasets.}
\label{ContourHub}
\end{figure}

\end{widetext}

\begin{widetext}

\begin{figure}[H]
\includegraphics[scale=0.6]{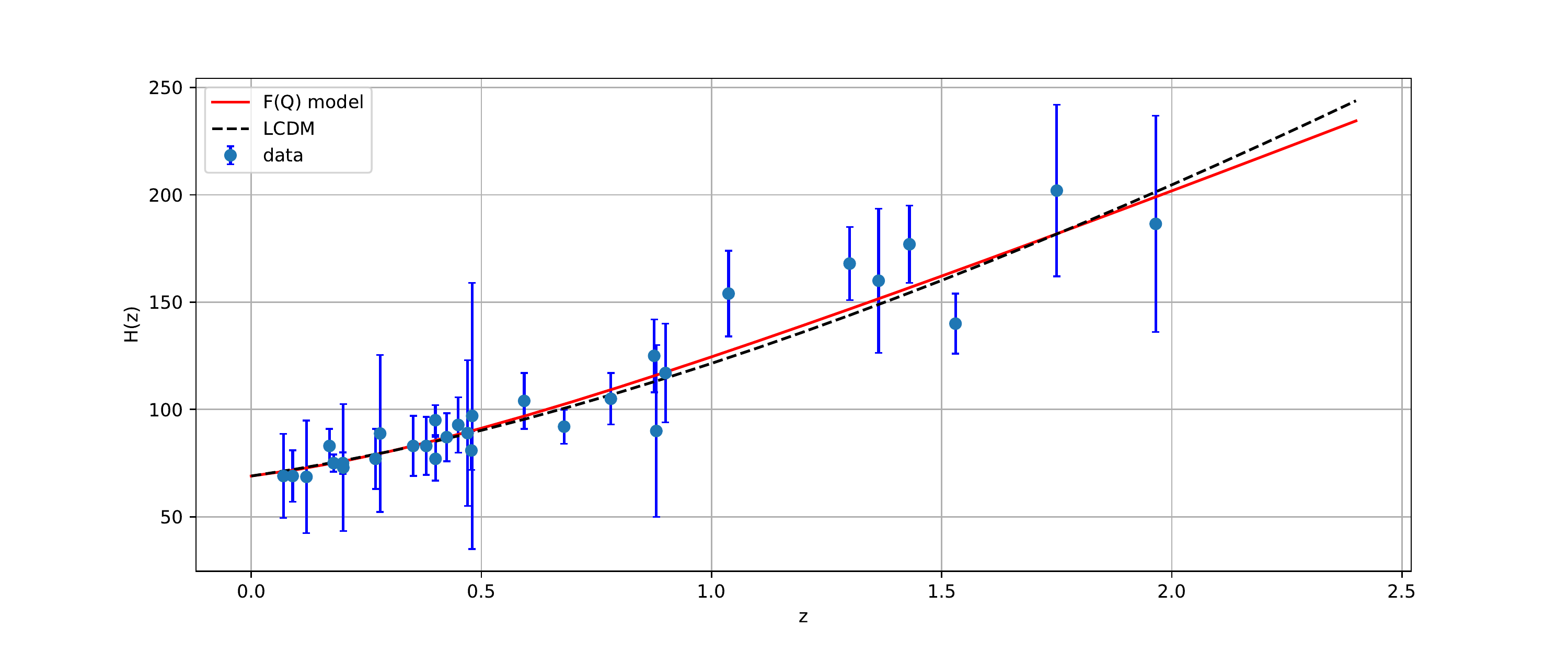}
\caption{The error bar plot of $H$ versus $z$ for the considered $f(Q)$ model. The solid red line is the curve for $f(Q)$ model whereas the black dotted line represents the $\Lambda$CDM model. The blue dots depict the 31 points of the Hubble data.} \label{ErrorHubble} 
\end{figure}
	
\end{widetext}

The best fit values of the model parameters are $\zeta= 0.65^{+0.10}_{-0.10} $, $\alpha=-0.0166^{+0.0079}_{-0.0053}$, $\gamma=1.36^{+0.14}_{-0.12}$ and $n=0.974^{+0.019}_{-0.027}$. 

Fig.\ref{ErrorHubble} shows the error bar plot of the considered model and $\Lambda$CDM or  standard cosmological model, with  cosmological constant density parameter $\Omega_{\Lambda_{0}}=0.7$, matter density parameter $\Omega_{m_{0}}=0.3$ and $H_{0}= 69$ km/s/Mpc. 

\subsection{BAO datasets}
The BAO distance dataset, which includes the 6dFGS, SDSS and WiggleZ surveys, comprise BAO measurements at six different redshifts in Table 1. The characteristic scale of BAO is ruled by the sound horizon $r_{s}$ at the epoch of photon decoupling $z_{\ast }$ that is given by the following relation:

\begin{equation}\label{4b}
r_{s}(z_{\ast })=\frac{c}{\sqrt{3}}\int_{0}^{\frac{1}{1+z_{\ast }}}\frac{da}{
a^{2}H(a)\sqrt{1+(3\Omega _{b0}/4\Omega _{\gamma 0})a}}.
\end{equation}
Here, $\Omega _{b0}$ and $\Omega _{\gamma 0}$ correspond to the present densities of baryons and photons respectively.

The following relations are used in BAO measurements

\begin{eqnarray}\label{4c}
 \triangle \theta &=&\frac{r_{s}}{d_{A}(z)},\\
\label{4d}
d_{A}(z) &=&\int_{0}^{z}\frac{dz^{\prime }}{H(z^{\prime })}, \\
	\triangle z &=&H(z)r_{s},
\end{eqnarray}
where $\triangle \theta $ represents the measured angular separation, $d_{A}$ is the angular diameter distance and $\triangle z$ represents the measured redshift separation of the BAO feature in the 2 point correlation function of the galaxy distribution on the sky along the line of sight.
   
In this work, BAO datasets of six points for $d_{A}(z_{\ast })/D_{V}(z_{BAO})$ is taken from the References \cite{BAO1, BAO2, BAO3, BAO4, BAO5, BAO6}, where the redshift at the epoch of photon decoupling is taken as $z_{\ast }\approx 1091$ and  $d_{A}(z)$ is the co-moving angular diameter
distance together with the dilation scale $D_{V}(z)=\left[
d_{A}(z)^{2}z/H(z)\right] ^{1/3}$. The chi-square function for the BAO datasets is taken to be \cite{BAO6} 
 
\begin{equation}\label{4e}
\chi _{BAO}^{2}=X^{T}C^{-1}X\,,  
\end{equation}

\begin{equation*}
X=\left( 
\begin{array}{c}
\frac{d_{A}(z_{\star })}{D_{V}(0.106)}-30.95 \\ 
\frac{d_{A}(z_{\star })}{D_{V}(0.2)}-17.55 \\ 
\frac{d_{A}(z_{\star })}{D_{V}(0.35)}-10.11 \\ 
\frac{d_{A}(z_{\star })}{D_{V}(0.44)}-8.44 \\ 
\frac{d_{A}(z_{\star })}{D_{V}(0.6)}-6.69 \\ 
\frac{d_{A}(z_{\star })}{D_{V}(0.73)}-5.45%
\end{array}%
\right) \,,
\end{equation*}

\begin{widetext}

\begin{table}[H]

\begin{center}

\begin{tabular}{|c|c|c|c|c|c|c|}
\hline
$z_{BAO}$ & $0.106$ & $0.2$ & $0.35$ & $0.44$ & $0.6$ & $0.73$ \\ \hline
$\frac{d_{A}(z_{\ast })}{D_{V}(z_{BAO})}$ & $30.95\pm 1.46$ & $17.55\pm 0.60$
& $10.11\pm 0.37$ & $8.44\pm 0.67$ & $6.69\pm 0.33$ & $5.45\pm 0.31$ \\ 
\hline
\end{tabular}
\caption{Values of $d_{A}(z_{\ast })/D_{V}(z_{BAO})$ for distinct values of $z_{BAO}$.}
\end{center}
\label{Table-1}
\end{table}

\end{widetext}

The inverse covariance matrix $C^{-1}$ is defined in
\cite{BAO6}

\begin{widetext}

\begin{equation*}
C^{-1}=\left( 
\begin{array}{cccccc}
0.48435 & -0.101383 & -0.164945 & -0.0305703 & -0.097874 & -0.106738 \\ 
-0.101383 & 3.2882 & -2.45497 & -0.0787898 & -0.252254 & -0.2751 \\ 
-0.164945 & -2.454987 & 9.55916 & -0.128187 & -0.410404 & -0.447574 \\ 
-0.0305703 & -0.0787898 & -0.128187 & 2.78728 & -2.75632 & 1.16437 \\ 
-0.097874 & -0.252254 & -0.410404 & -2.75632 & 14.9245 & -7.32441 \\ 
-0.106738 & -0.2751 & -0.447574 & 1.16437 & -7.32441 & 14.5022%
\end{array}%
\right) \,.
\end{equation*}

\end{widetext}

\begin{widetext}

\begin{figure}[H]
\centering
\includegraphics[scale=0.9]{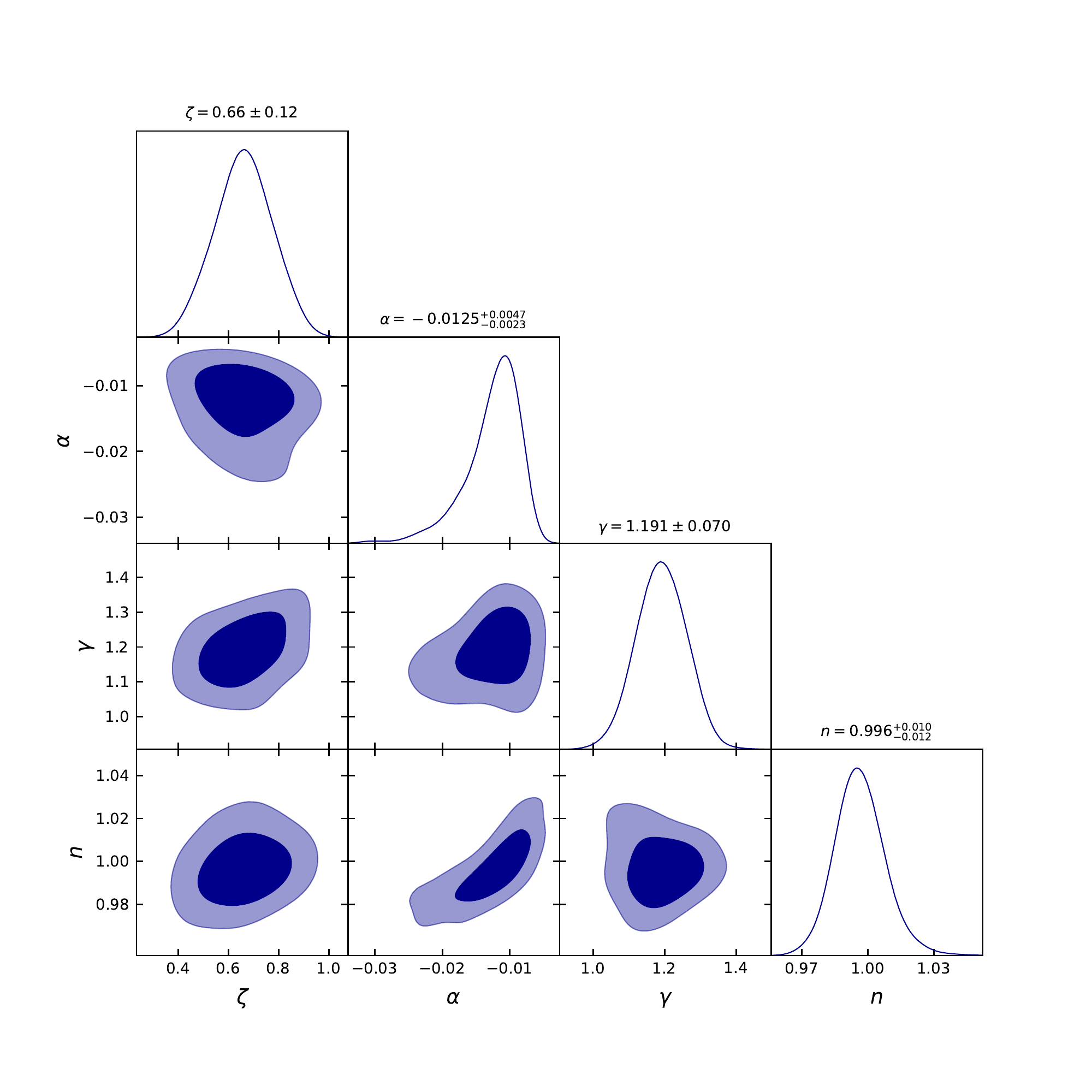}
\caption{The $1-\sigma$ and $2-\sigma$ likelihood contours for the model parameters using the BAO datasets.}
\label{ContourBAO}
\end{figure}	

\end{widetext}

The values that fit observations are } $\zeta=0.66^{+0.12}_{-0.12}$, $\alpha=-0.0125^{+0.0047}_{-0.0023}$, $\gamma=1.191^{+0.070}_{-0.070}$ and $n=0.996^{+0.010}_{-0.012}$ in Fig.\ref{ContourBAO}.

\subsection{Pantheon datasets} 

Scolnic et al. \cite{Scolnic/2018} put together the Pantheon samples consisting of 1048 type Ia supernovae in the redshift range $0.01 < z < 2.3$. The PanSTARSS1 Medium Deep Survey, SDSS, SNLS and numerous low-z and HST samples contribute to it. The empirical relation used to calculate the distance modulus of SNeIa from the observation of light curves is given by $\mu= m_{B}^{*}+\alpha X_{1}-\beta C-M_{B} + \Delta_{M}+ \Delta_{B}$, where $X_{1}$ and $C$ denote the stretch and color correction parameters, respectively \cite{Scolnic/2018,Mukherjee/2021}, $m_{B}^*$ represents the observed apparent magnitude and $M_{B}$ is the absolute magnitude in the B-band for SNeIa. The parameters $\alpha$ and $\beta$ are the two nuisance parameters describing the luminosity stretch and luminosity color relations, respectively. Further, the distance correction factor is $\Delta_{M}$ and $\Delta_{B}$ is a distance correction based on predicted biases from simulations.

The nuisance parameters in the Tripp formula \cite{Tripp/1998} were reconstructed using a novel technique called BEAMS with Bias Corrections \cite{Kessler/2017, Fotios/2021} and the observed distance modulus was reduced to the difference between the corrected apparent magnitude $m_{B}$ and the absolute magnitude $M_{B}$, which is $\mu= m_{B}-M_{B}$. We shall avoid marginalizing the over nuisance parameters $\alpha$ and $\beta$ but marginalize over the Pantheon data for $M_{B}$. Hence, we ignore the values of $\alpha$ and $\beta$ for the present investigation of the model.

The luminosity distance read as

\begin{equation}\label{4f}
D_{L}(z)= (1+z) \int_{0}^{z} \frac{c dz'}{H(z')},
\end{equation}
 with $c$ being the speed of light.

The $\chi^{2}$ function for type Ia supernovae is obtained by correlating the theoretical distance modulus 
\begin{eqnarray}\label{4g}
\mu(z) &=& 5log_{10}D_{L}(z)+\mu_{0},\\
\label{4h}
\mu_{0} &=&  5log(1/H_{0}Mpc) + 25,
\end{eqnarray}
such that

\begin{equation}\label{4i}
\chi^{2}_{SN}(\zeta, \alpha, \gamma, n)= \sum _{k=1}^{1048} \dfrac{\left[ \mu_{obs}(z_{k})-\mu_{th}(\zeta, \alpha, \gamma, n, z_{k})\right] ^{2}}{\sigma^{2}(z_{k})},
\end{equation}
where $\mu_{th}$ is the theoretical value of distance modulus,  $\mu_{obs}$ is the observed value whereas $\sigma^{2}(z_{k})$ is the standard error in the observed value.

\begin{widetext}

\begin{figure}[H]
\centering
\includegraphics[scale=0.9]{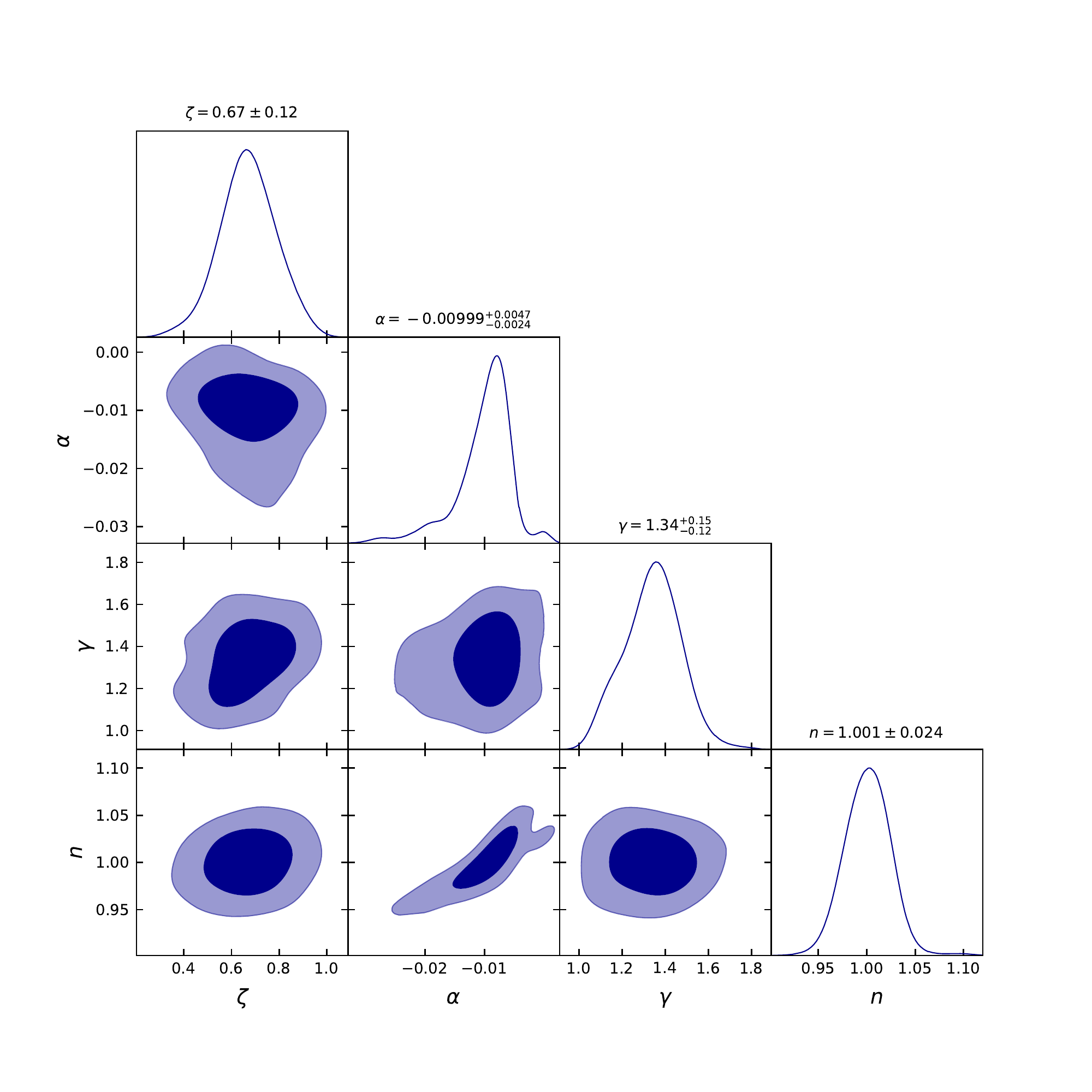}
\caption{The $1-\sigma$ and $2-\sigma$ likelihood contours for the model parameters using the Pantheon datasets.}
\label{ContourPan}
\end{figure}	

\end{widetext}

Using the Pantheon supernovae datasets, we obtain the best fit values for $\zeta$, $\alpha$, $\gamma$ and $n$ as the $1-\sigma$ and $2-\sigma$ contour plots in Fig.\ref{ContourPan}. The values that fit the model are $\zeta=0.67^{+0.12}_{-0.12} $, $\alpha=-0.00999^{+0.0047}_{-0.0024}$, $\gamma=1.34^{+0.15}_{-0.12}$ and $n=1.001^{+0.024}_{-0.024}$. 

\subsection{Cosmological Parameters}

The evolution of the Hubble parameter, deceleration parameter, energy density, pressure with bulk viscosity and the effective EoS parameter for the redshift range $-1 < z <8$ are presented below, in order to test the late time cosmic expansion history and the future of expanding universe \cite{Sunny}. In order to do so we use the set of values constrained by Hubble, BAO and the Pantheon data sets for the model parameters.

\begin{figure}[H]
\includegraphics[scale=0.51]{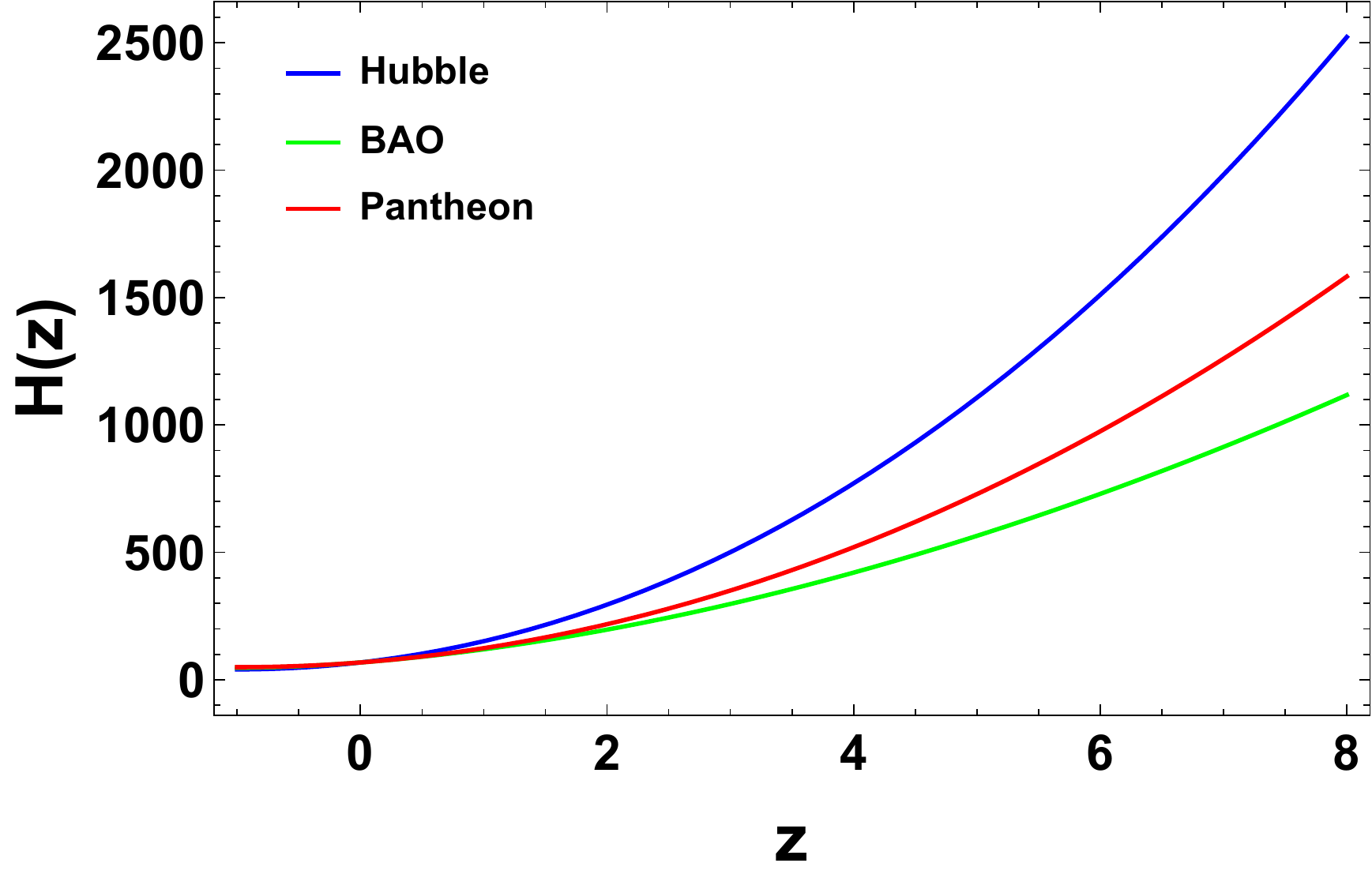}
\caption{Profile of the Hubble parameter for the given model corresponding to the values of the parameters constrained by the Hubble, BAO and the Pantheon data point sets.}
\label{h}
\end{figure}

\begin{figure}[H]
\includegraphics[scale=0.5]{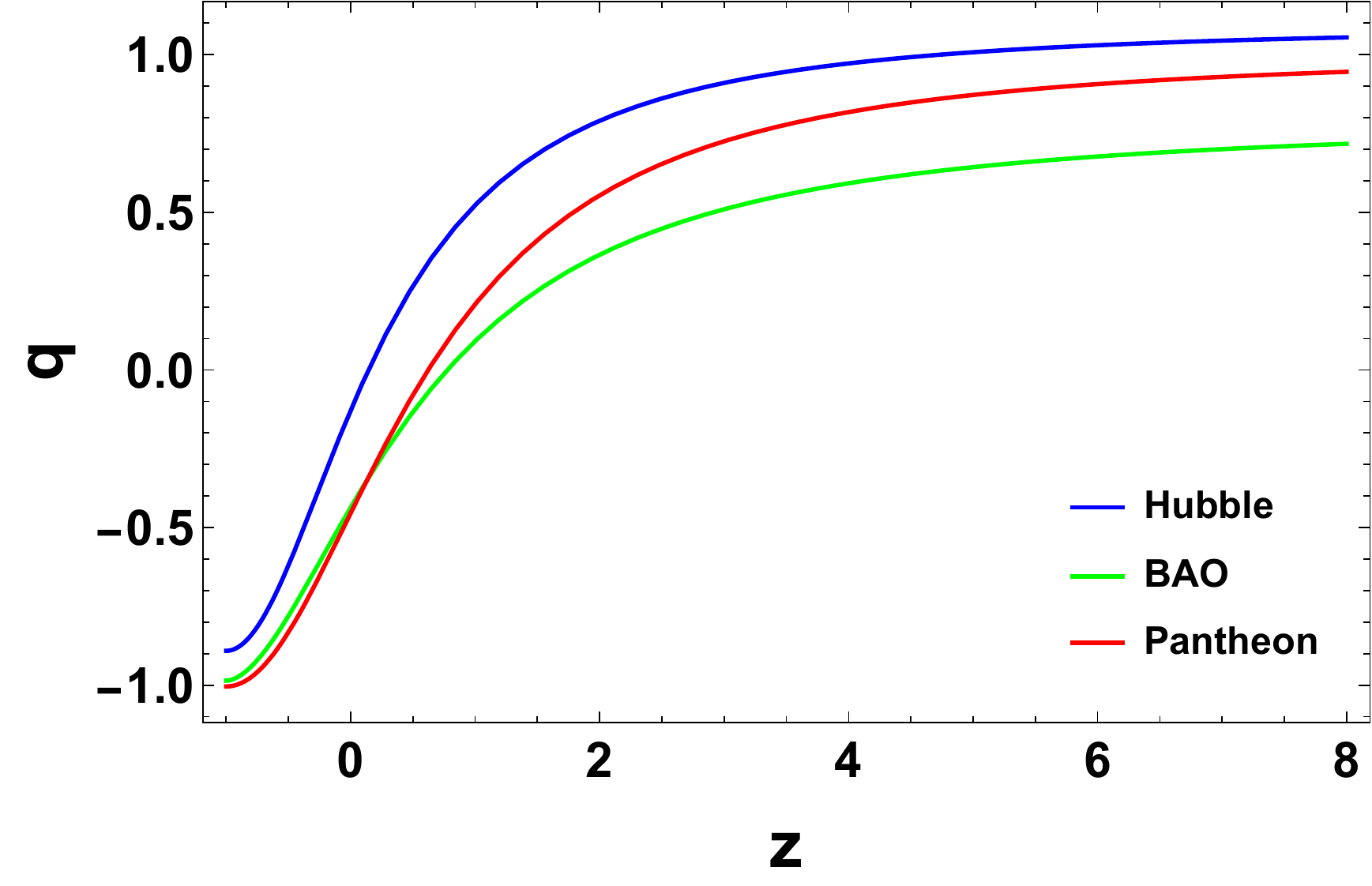}
\caption{Profile of the deceleration parameter for the given model corresponding to the values of the parameters constrained by the Hubble, BAO and the Pantheon data point sets.}
\label{q}
\end{figure}

\begin{figure}[]
\centering
\includegraphics[scale=0.52]{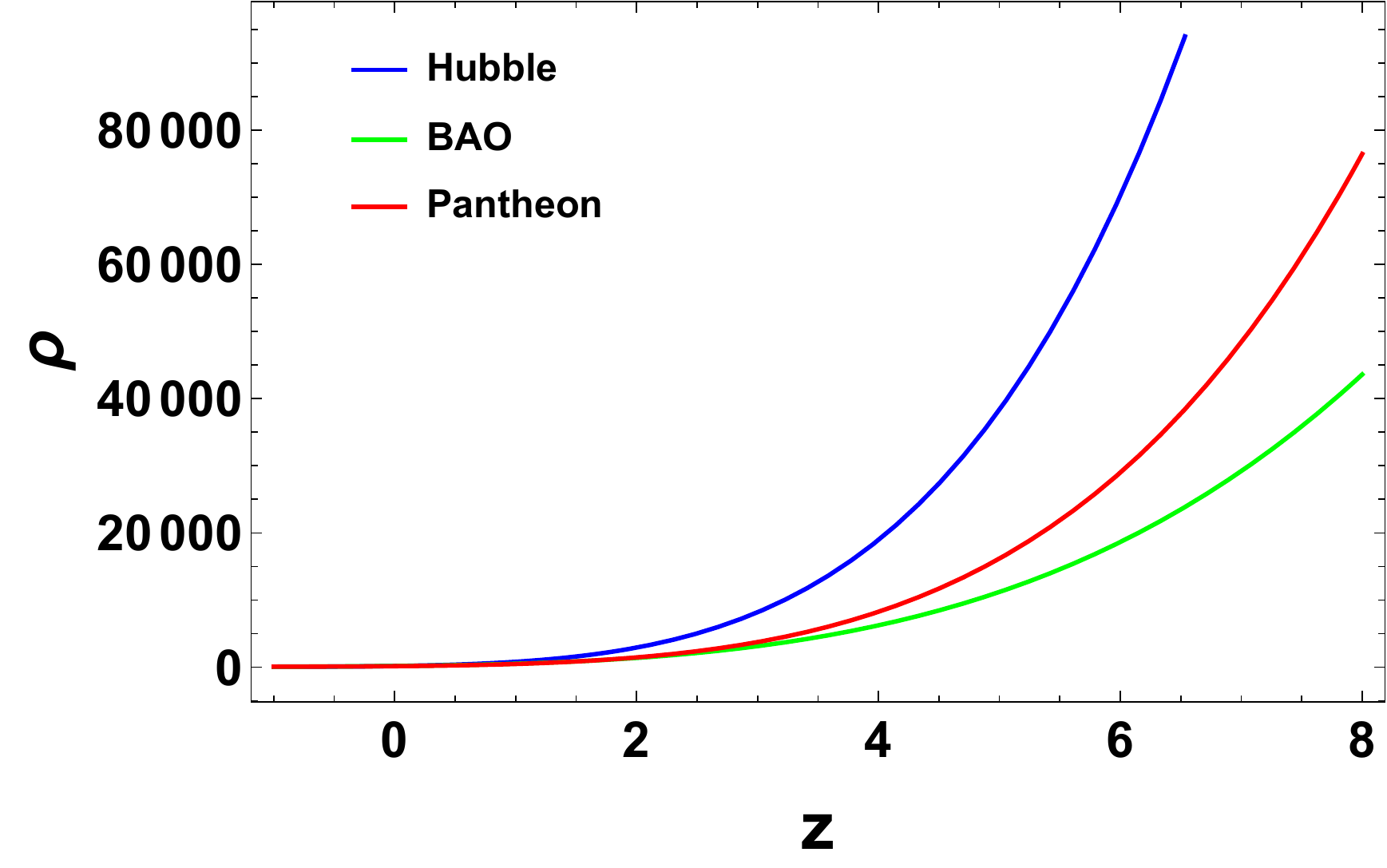}
\caption{Profile of the density parameter for the given model corresponding to the values of the parameters constrained by the Hubble, BAO and the Pantheon data point sets.}
\label{rho}
\end{figure}

\begin{figure}[]
\centering
\includegraphics[scale=0.41]{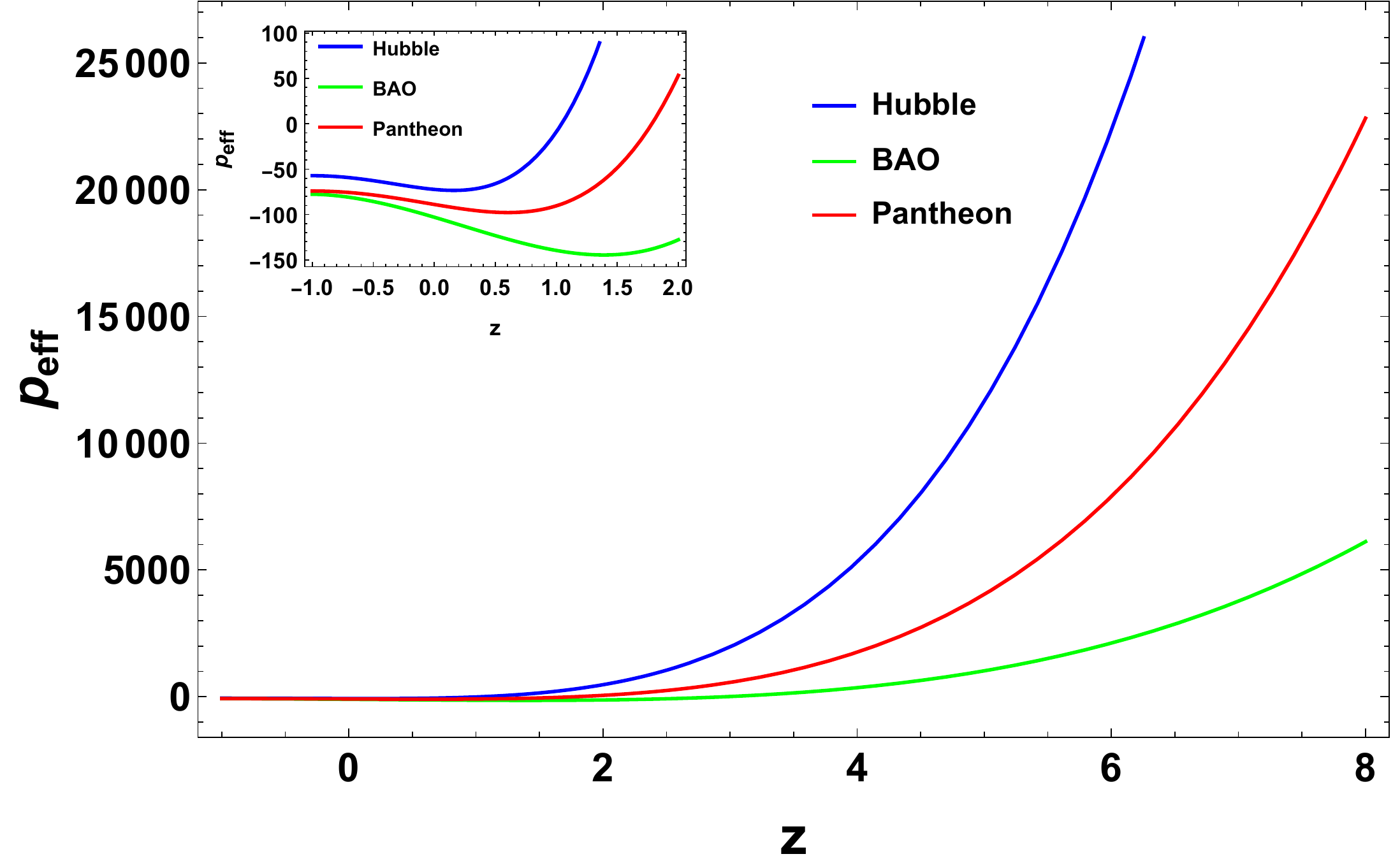}
\caption{Profile of the pressure for the given model corresponding to the values of the parameters constrained by the Hubble, BAO and the Pantheon data point sets.}
\label{p}
\end{figure}

\begin{figure}[]
\centering
\includegraphics[scale=0.5]{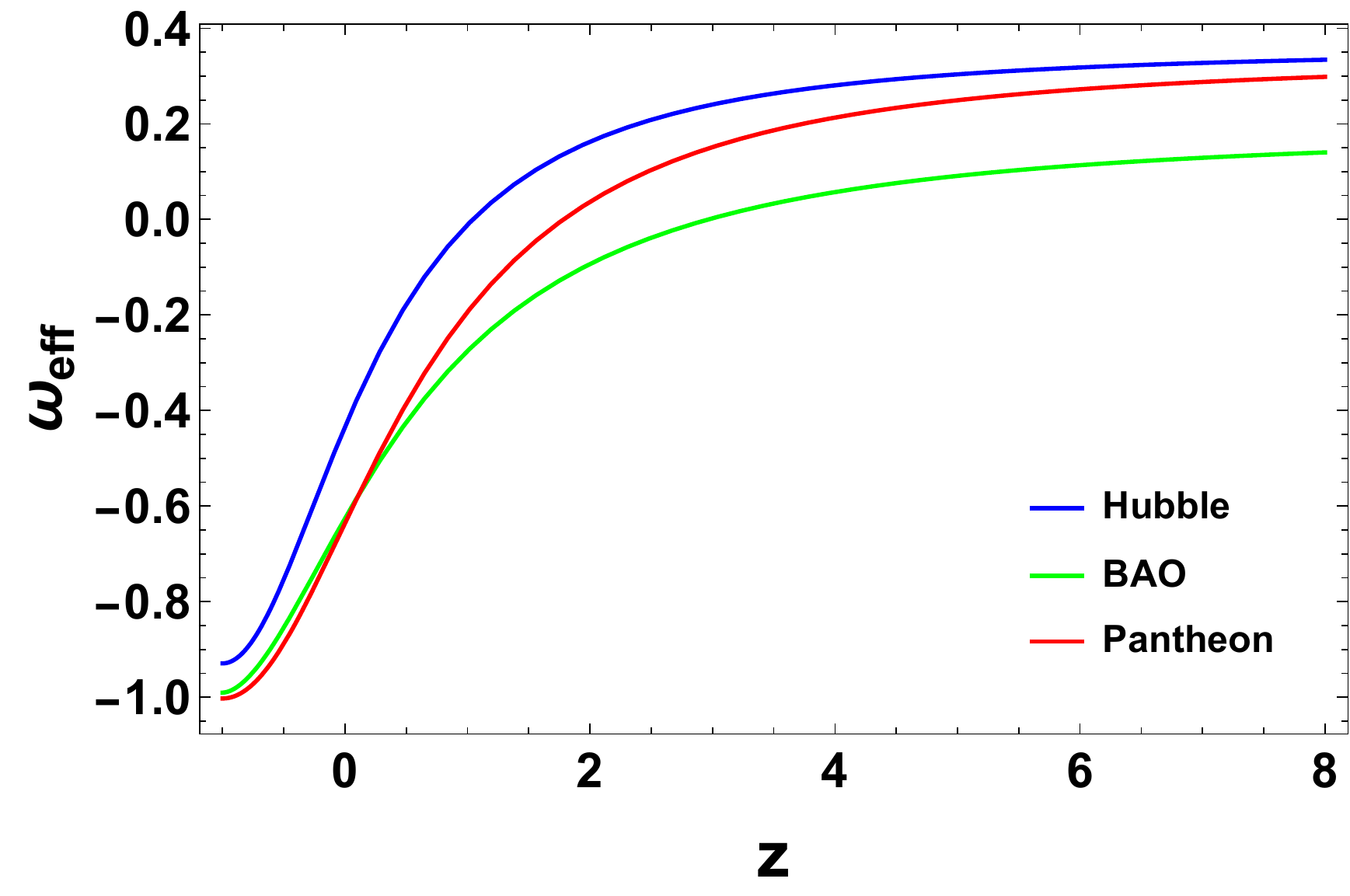}
\caption{Profile of the EoS parameter for the given model corresponding to the values of the parameters constrained by the Hubble, BAO and the Pantheon data point sets.}
\label{w}
\end{figure}

From Fig.\ref{q}, it is clear that the deceleration parameter shows the transition from a decelerated ($q>0$) to an accelerated ($q<0$) phase of the universe expansion for the constrained values of the model parameters. The transition redshift is $z_t \thickapprox 0.142$,   $ z_t \thickapprox 0.776 $ and $ z_t \thickapprox 0.622 $ corresponding to the Hubble, BAO and the Pantheon data sets, respectively. The present value of the deceleration parameter is, respectively,  $q_0 = -0.127$ , $q_0= -0.436$ and  $q_0= -0.454$. 

From Fig. \ref{h} and \ref{rho} it is clear that the Hubble and density parameter shows the positive behavior for all the constrained values of the model parameters, which is expected.

Fig \ref{p} indicates that the bulk viscous cosmic fluid exhibits, for lower redshifts, the negative pressure that make bulk viscosity to be a viable candidate to drive the cosmic acceleration. Furthermore, the effective EoS parameter presented in Fig.\ref{w} indicates that the cosmic viscous fluid behaves like quintessence dark energy. The present values of EoS parameter corresponding to the Hubble, BAO and the Pantheon samples are  $\omega_{0}= -0.433$ , $\omega_0= -0.625$, and  $\omega_0= -0.635$. 
%

\section{Energy Conditions}\label{sec5}

In the present section we are going to construct the energy conditions for the solutions of the present model. The energy conditions are relations applied to the matter energy-momentum tensor with the purpose of satisfying positive energy. The energy conditions are derived from the Raychaudhuri equation and are written as \cite{EC} \\

\begin{itemize}
\item \textbf{Null energy condition (NEC) :} $\rho_{eff}+p_{eff}\geq 0$;  
\item \textbf{Weak energy condition (WEC) :} $\rho_{eff} \geq 0$ and  $\rho_{eff}+p_{eff}\geq 0$; 
\item \textbf{Dominant energy condition (DEC) :} $\rho_{eff} \pm p_{eff}\geq 0$; 
\item \textbf{Strong energy condition (SEC) :} $\rho_{eff}+ 3p_{eff}\geq 0$,
\end{itemize}
 with $\rho_{eff}$ being the effective energy density.

\begin{figure}[H]
\centering
\includegraphics[scale=0.52]{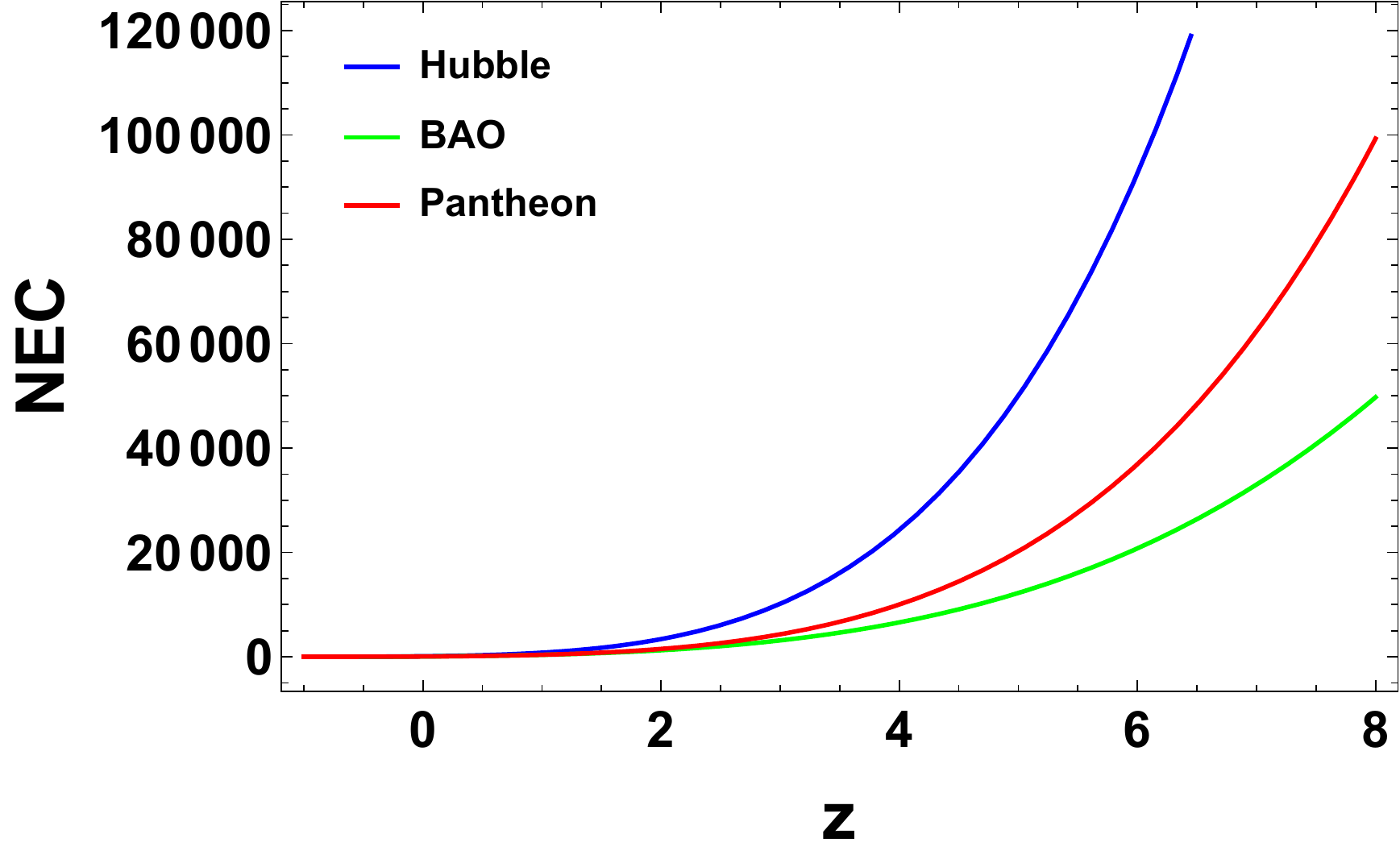}
\caption{Profile of the null energy condition for the given model corresponding to the values of the parameters constrained by the Hubble, BAO and the Pantheon data point sets.}
\label{nec}
\end{figure}

\begin{figure}[H]
\centering
\includegraphics[scale=0.51]{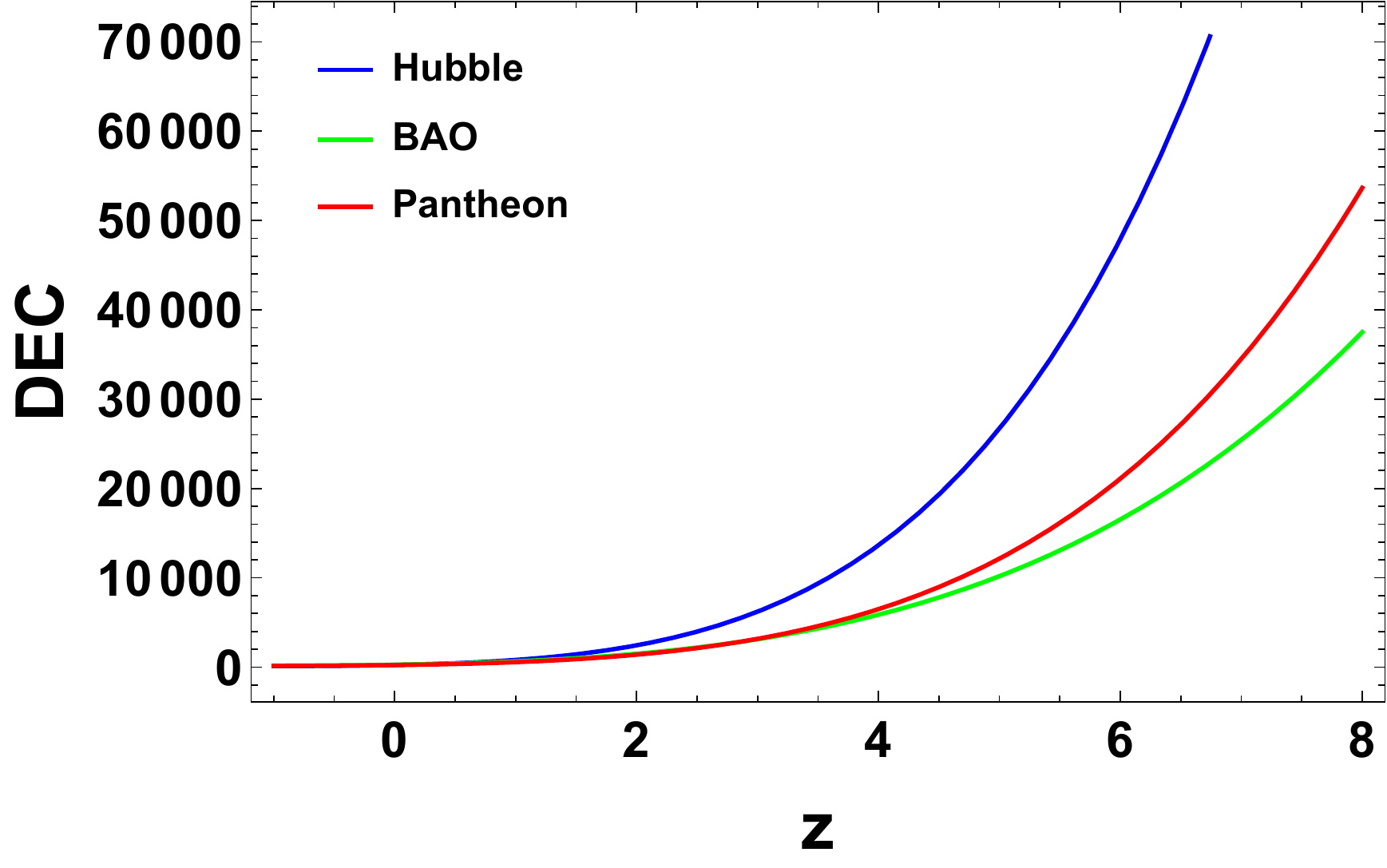}
\caption{Profile of the dominant energy condition for the given model corresponding to the values of the parameters constrained by the Hubble, BAO and the Pantheon data point sets.}
\label{dec}
\end{figure}

\begin{figure}[H]
\centering
\includegraphics[scale=0.45]{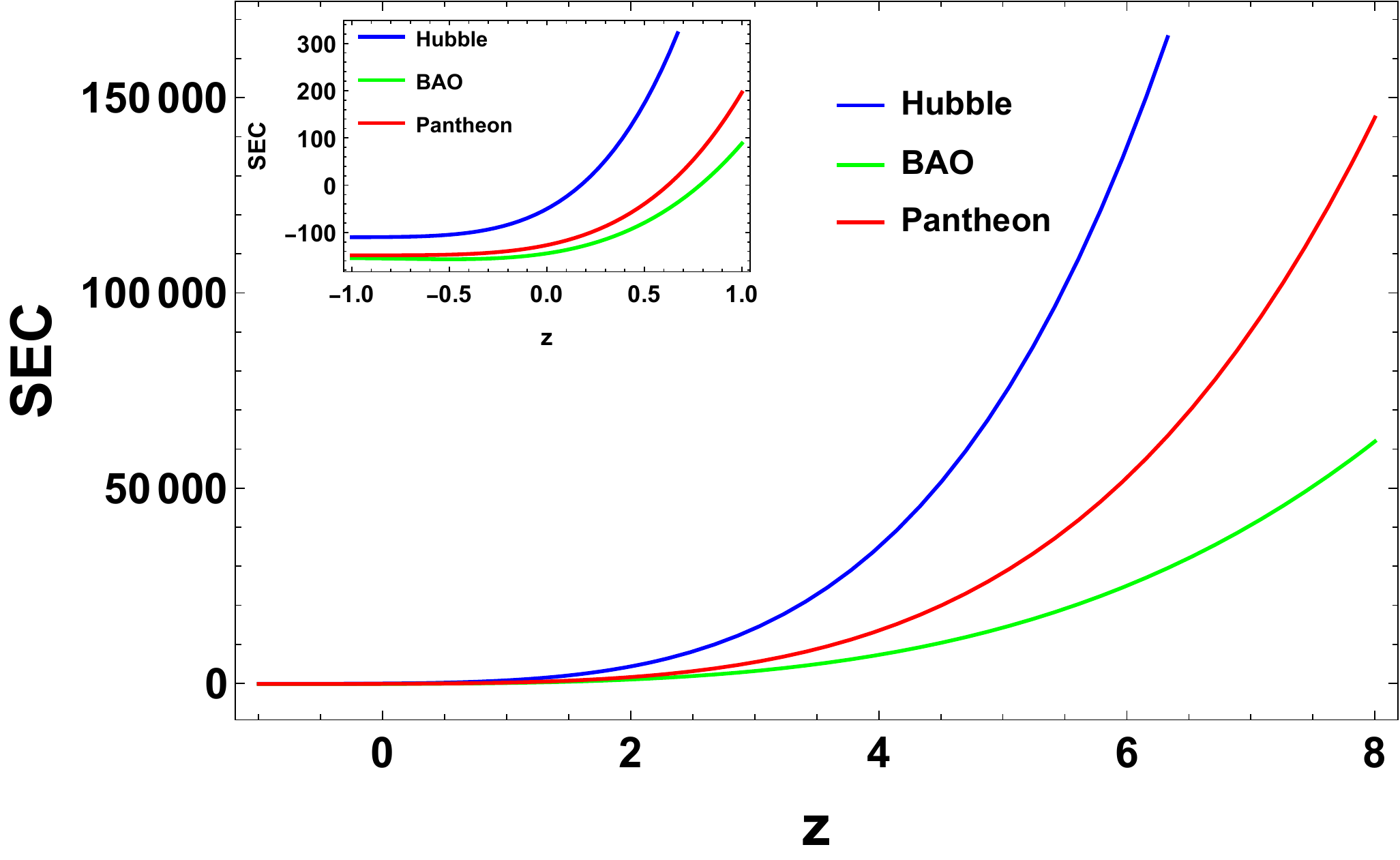}
\caption{Profile of the strong energy condition for the given model corresponding to the values of the parameters constrained by the Hubble, BAO and the Pantheon data point sets.}
\label{sec}
\end{figure}

In Figs.\ref{nec} and \ref{dec} it is evident that the NEC and DEC exhibit positive behavior for all the constrained values of the model parameters. As WEC is the combination of energy density and NEC, we conclude that NEC, DEC and WEC are all satisfied in the entire domain of redshift. Fig.\ref{sec} indicates that the SEC exhibits, for lower redshifts, the negative behavior that is related to cosmic acceleration \cite{Mandal/2020}. This is also reflected in the deceleration parameter behavior in Fig.\ref{q}.

\section{Statefinder analysis}\label{sec6}

The cosmological constant $\Lambda$ suffers from two major drawbacks, namely the aforementioned cosmological constant and cosmic coincidence problems. To surmount these problems, dynamic models of dark energy have been introduced in the literature as we have also mentioned in Introduction. To discriminate between these time-varying dark energy  models, an appropriate tool was required. In this direction, V. Sahni et al. introduced a new pair of geometrical parameters known as {\it statefinder parameters} $(r,s)$ \cite{VS}. The statefinder parameters are defined as
\begin{eqnarray}
r &=&\frac{\dddot{a}}{aH^3},\\
s&=&\frac{(r-1)}{3(q-\frac{1}{2})}.
\end{eqnarray}
The parameter $r$ can be rewritten as $r=2q^2+q-\frac{\dot{q}}{H}$.

For different values of the statefinder pair $(r,s)$, it represents the following dark energy models:
\\ \\
${r=1,s=0}$ represents $\Lambda$CDM model,\\
${r>1,s<0}$ represents Chaplygin gas model,\\
${r<1,s>0}$ represents quintessence model.\\

\begin{figure}[]
\includegraphics[scale=0.5]{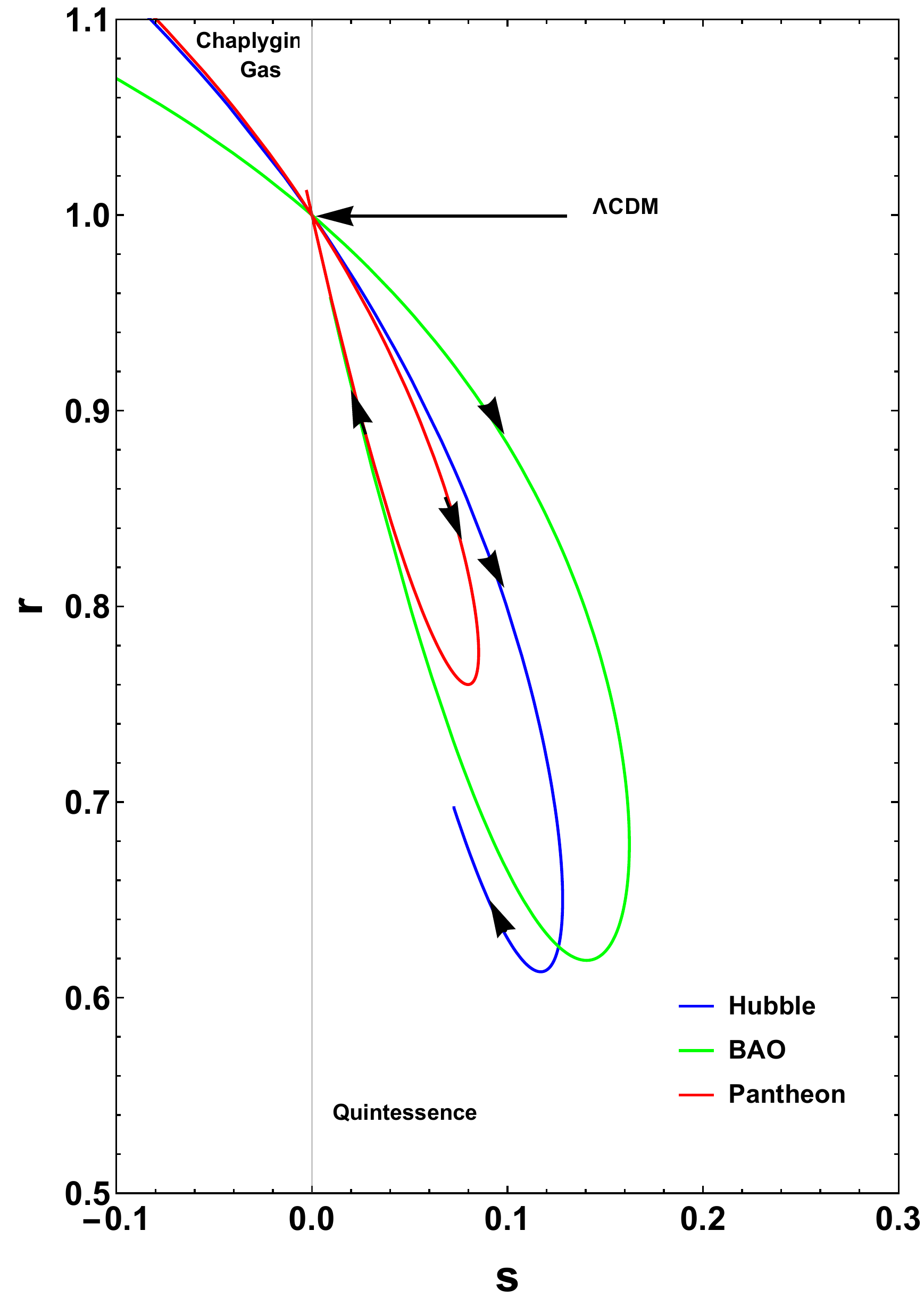}
\caption{Plot of the trajectories in the $r-s$ plane for the given cosmological model corresponding to the parameters values constrained by the Hubble, BAO and the Pantheon data sets.}
\label{rs}
\end{figure}

\begin{figure}[]
\includegraphics[scale=0.5]{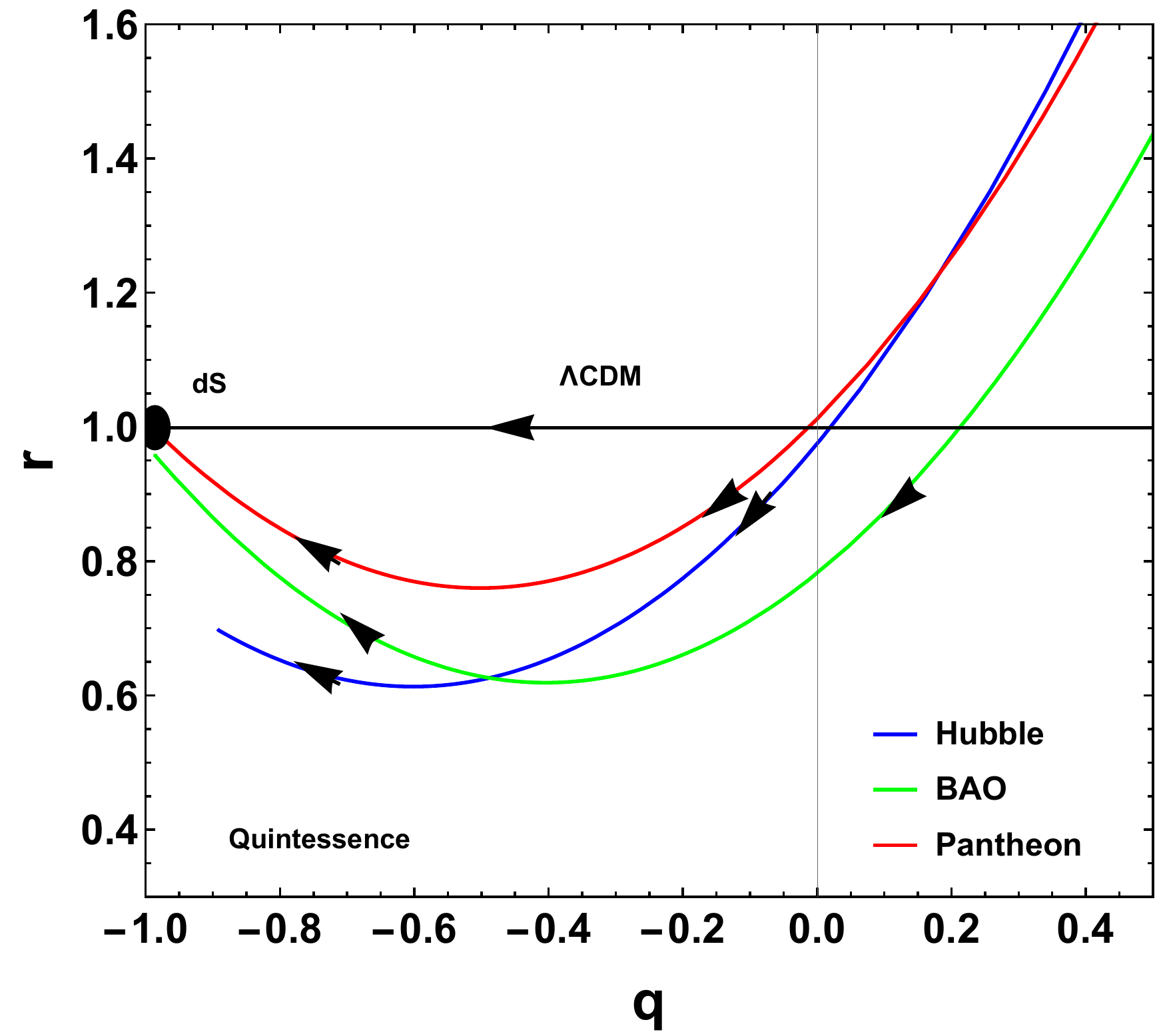}
\caption{Plot of the trajectories in the $q-r$ plane for the given cosmological model corresponding to the parameters values constrained by the Hubble, BAO, and the Pantheon data sets.}
\label{qr}
\end{figure}

In Figures \ref{rs} and \ref{qr}, we plot the $s-r$ and $q-r$ diagrams for our cosmological model by taking the values of the parameters constrained by the Hubble, BAO and the Pantheon data sets. 

Figures \ref{rs} and \ref{qr} show that our bulk viscous model lies in the quintessence region. Also, the evolutionary trajectories of our model departure from the $\Lambda$CDM point. The present values of the statefinder parameters corresponding to the values of the model parameters constrained by the Hubble, BAO and the Pantheon samples are 
$ r_0= 0.837$ and $s_0= 0.086$, $r_0= 0.62$ and $s_0= 0.135$,  $r_0= 0.762$ and $s_0=  0.083$ respectively.

\section{Discussions and conclusions}\label{sec7}

In the present section we will discuss the results obtained in Sections IV,V, and VI for the bulk viscous symmetric teleparallel cosmological model here developed and presented.

Cosmology has been on the agenda mainly for two reasons: dark energy and dark matter. While dark energy has been deeply discussed throughout the paper, dark matter is predicted within $\Lambda$CDM model as a sort of matter that does not interact electromagnetically, so that it cannot be seen, but its gravitational effects otherwise can well be detected. Still we have not yet detected or even associated dark matter to a particle of standard model or beyond \cite{cdmsii_collaboration/2010,akerib/2014,essig/2012}. Modified (or alternative) theories of gravity have also been used to describe dark matter effects \cite{bohmer/2008,mannheim/2012}. In these cases, dark matter is simply an effect of modification of gravity.

Returning to the dark energy question, it is highly counter-intuitive that the expansion of the universe is actually accelerating. Although the vacuum quantum energy can well explain this dynamical effect via the cosmological constant in Einstein's field equations of GR, the aforementioned important and persistent problems related to $\Lambda$ supply the search for alternative explanations.

In the present article, as an attempt to describe dark energy we have assumed the symmetric teleparallel gravity as the underlying gravity theory.

The $f(Q)$ gravity was recently proposed by Jim\'enez et al. in \cite{J.B.} as a ramification of the geometric trinity, that says that space-time manifold can be described by curvature, torsion or non-metricity. Particularly, the symmetric teleparallel gravity describes gravitational interactions via the non-metricity scalar, with null curvature and torsion.

Our $f(Q)$ cosmological model was based on a spatially homogeneous and isotropic flat metric and an energy-momentum tensor describing a bulk viscous fluid. We let the $f(Q)$ function be a power of $n$ as $f(Q)\sim Q^n$, with $n$ a free parameter.

In Section IV we started testing our cosmological solutions. We started confronting the Hubble parameter \eqref{3m} with 31 data points that are measured from differential age approach. In Fig.\ref{ContourHub} we have obtained the best fit values for the model free parameters. We have then plotted Fig.\ref{ErrorHubble}, in which our $H(z)$ is confronted with cosmological data and compared with $\Lambda$CDM prediction. We can see the $f(Q)$ model describes observations with good agreement and specially for higher values of redshift it is clear that it provides a better fit when compared to $\Lambda$CDM model. Further, in Figs.\eqref{ContourBAO} and \eqref{ContourPan}, we have obtained the best fit values for the model free parameters.

From Fig. \ref{h} and \ref{rho} we found that the Hubble and density parameter shows the expected positive behavior for all the constrained values of the model parameters.
Fig.\ref{p} indicates that the bulk viscous cosmic fluid exhibit the negative pressure that make bulk viscosity to be a viable candidate to drive the cosmic acceleration. This is also reflected in the deceleration parameter behavior in Fig.\ref{q}, which shows a transition from decelerated to accelerated phases of the universe expansion. Furthermore, the effective EoS parameter presented in Fig.\ref{w} indicates that the cosmic viscous fluid behaves like quintessence dark energy.

In Section V, we investigated the consistency of our model by analyzing the different energy conditions. We found that NEC, DEC, and WEC all are satisfied in the entire domain of redshift (presented in Fig\ref{nec} and \ref{dec}) while the SEC, presented in Fig\ref{sec}, is violated for lower redshifts that implies the cosmic acceleration and satisfied for higher redshifts that implies a decelerated phase of the universe.

In Section VI, Figs.\ref{rs} and \ref{qr} show that the evolutionary trajectories of our model is departed from $\Lambda$CDM fixed point {${r=1,s=0}$}. In the present epoch they lie in the quintessence region {${r<1,s>0}$}. The present model is, therefore, a good alternative to explain the universe dynamics, particularly with no necessity of invoking the cosmological constant. 

 Finally, it is worth remarking that $f(Q)$ gravity still needs to be applied in several different regimes in order to be stablished as a viable gravitational formalism. In \cite{J.B.} it was shown that the $f(Q)$ gravity can be compatible with Solar System constraints. Recently, in \cite{frusciante/2021} it was shown that the $f(Q)$ gravity predicts a modified gravitational wave propagation and a different gravitational wave luminosity distance when compared to the standard electromagnetic one, although this investigation was done for a different functional form of the $f(Q)$ function. \\
 

  
\section*{ACKNOWLEDGMENTS}

RS acknowledges University Grants Commission (UGC), New Delhi, India for awarding Junior Research Fellowship (UGC-Ref. No.: 191620096030). SA  acknowledges  CSIR,  New  Delhi,  India  for  SRF. PKS   acknowledges   CSIR,   New   Delhi,   India   for   financial   support   to carry   out   the   Research   project [No.03(1454)/19/EMR-II Dt.02/08/2019] and IUCAA, Pune, India for providing support through visiting associateship program.

\end{document}